\documentstyle{article}
\begin{document}

\title{The Solar Neutrino Problem - An Update} 
\author{Arnon Dar and Giora Shaviv\\
{\it Department of Physics and Space Research Institute}\\
{\it Technion-Israel Institute of Technology, Haifa 32000, Israel}\\}
\date{}
\maketitle

\begin{abstract} 

The $^8$B solar neutrino flux that has recently been measured by
Super-Kamiokande is consistent with the $^{37}$Ar production rate in
$^{37}$Cl at Homestake. The gallium solar neutrino experiments, GALLEX and
SAGE, continue to observe $^{71}$Ge production rates in $^{71}$Ga that are
consistent with the minimal signal expected from the solar luminosity. The
observed $^8$B solar neutrino flux is in good agreement with that
predicted by the standard solar model of Dar and Shaviv with nuclear
reaction rates that are supported by recent measurements of nuclear fusion
cross sections at low energies.  But, the signals measured by
Super-Kamiokande, SAGE and GALLEX leave no room for the contribution from
the expected $^7$Be solar neutrino flux. This apparent suppression of the
$^7$Be solar neutrino flux can be explained by neutrino oscillations and
the Mikheyev-Smirnov-Wolfenstein effect, although neither a flavor change,
nor a terrestrial variation, nor a spectral distortion of the $^8$B solar
neutrino flux has been observed. Detailed helioseismology data from
SOHO and GONG confirm the standard solar model description of the solar
core, but helioseismology is insensitive to the fate of $^7$Be in the sun
and the production rates of $^7$Be, CNO and $^8$B neutrinos. Thus,
helioseismology and the solar neutrino problem do not provide conclusive
evidence for neutrino properties beyond the standard electroweak model. 
The solar neutrino problem may still be an astrophysical problem. The
deviations of the experimental results from those predicted by the
standard solar models may reflect the approximate nature of our knowledge
of nuclear reaction rates and radiation transport in dense stellar plasmas
and the approximate nature of solar models.  Only future observations of
spectral distortions, or terrestrial modulation or flavor change of solar
neutrinos in solar neutrino experiments, such as Super-Kamiokande, SNO,
Borexino and HELLAZ will be able to establish that neutrino properties
beyond the minimal standard electroweak model are responsible for the
solar neutrino problem. 

\end{abstract}
\leftline{Keywords: Sun; Standard Solar Model; Neutrinos; Helioseismology;}

\section{Introduction} 

The sun is a typical main sequence star that generates its energy via
fusion of hydrogen into helium through the pp and CNO nuclear reaction
chains (Fig. 1). Due to the conservation laws of baryon number, electric
charge, lepton flavor and energy, the total solar neutrino flux is fixed,
practically, by the solar luminosity (e.g., Dar and Nussinov 1991). The
neutrino spectrum is essentially a sum of standard beta decay spectra from
the $\beta$-decays ${\rm 2p \rightarrow De^+\nu_e}$, $^8$ B$\rightarrow 
2\alpha e^+\nu_e$, $^{13}$N$\rightarrow ^{13}$C$e^+\nu_e$ and
$^{15}$O$\rightarrow^{15}$N$e^+\nu_e$ and ``lines'' from the electron
captures $e^7$Be$\rightarrow\nu_e ^7$Li and ${\rm pep\rightarrow D\nu_e}$. 
To a good
approximation, they are independent of the conditions in the sun. However,
the relative contributions of the various solar neutrino sources depend on
the chemical composition, temperature and density distributions near the
center of the sun. These are usually estimated from standard solar models
(SSM). They can be tested also by helioseismology, which, however, is not
sensitive to the exact abundance or fate of trace elements (e.g., $^7$Be)
in the sun. 

Solar neutrinos, have been detected on Earth in roughly their expected
numbers, in five underground solar neutrino experiments, the Chlorine
solar neutrino experiment at Homestake, South Dakota, USA, the Water
Cherenkov Experiment, Kamiokande, at Kamioka Japan, the Soviet-American
Gallium Experiment, SAGE, at the Baksan, Russia, the European Gallium
Experiment, GALLEX, at Gran Sasso, Italy and the large water Cherenkov
Experiment, Super-Kamiokande, at Kamioka Japan. These experiments have
confirmed that the sun is powered by fusion of hydrogen into helium. This
milestone achievement in physics, however, has been overshadowed by the
fact that the combined results from the solar neutrino experiments seem to
suggest that the solar neutrino flux differs significantly from that
expected from the standard solar models. This discrepancy has become known
as the solar neutrino problem.  Many authors have argued that the solar
neutrino problem provides conclusive evidence for neutrino properties
beyond the minimal standard electroweak model. However, conclusive
evidence for new electroweak physics from solar neutrino observations can
be provided only by detecting at least one of the following signals: 

\noindent
1. Solar neutrinos with flavors other than $\nu_e$.         

\noindent
2. Spectral distortion of the fundamental $\beta$-decay spectra.

\noindent 
3. Terrestrial modulations of the solar neutrino flux.

\noindent
4. A clear violation of the luminosity sum rule.

\noindent 
So far, no such conclusive evidence has been provided by the
solar neutrino experiments. Therefore, the solar neutrino problem does not
provide solid evidence for neutrino properties beyond the standard
electroweak model and standard physics solutions to the solar neutrino
problem are not ruled out. Moreover, a closer look at the sun through
helioseismology, X-ray and UV observations shows that the sun is a
bewildering turmoil of complex phenomena. It shows unexpected features and
behavior at any scale. It has a strange complex internal rotation,
unexplained magnetic activity with unexplained 11 year cycle, unexpected
anomalies in its surface elemental abundances, unexplained explosions in
its atmosphere and unexplained mechanism that heats its million degree
corona and accelerates the solar wind.  Perhaps the surface of the sun is
complex because we can see it and the center of the sun is not only
because we cannot?  Perhaps the SSM which has been improved continuously
over the past three decades but which still uses simple plasma physics and
assumes an exact spherical symmetry, no mass loss or mass accretion, no
angular momentum loss or gain, no differential rotation, zero magnetic
field through the entire solar evolution, is a too simplistic picture and
does not provide a sufficiently accurate description of the core of the
sun. 

In this paper we summarize the experimental results from the various solar
neutrino experiments, we discuss shortly the solar neutrino problem in a
``model independent way'' and we compare the experimental results with
updated standard solar model calculations. We conclude that (a) there is
no conclusive evidence for a $^8$B solar neutrino problem, (b) the $^8$B
solar neutrino flux as measured by Super-Kamiokande is in good agreement
with that predicted by the standard solar model (of Dar and Shaviv) with
nuclear reaction rates that are supported by recent measurements of
nuclear reaction rates at low energies, (c) the suppression of $^7$Be
solar neutrinos which is suggested by both the chlorine and gallium
experiments can be due to neutrino oscillations and the
Mikheyev-Smirnov-Wolfenstein effect, although neither a day-night effect,
nor a spectral distortion of the $^8$B solar neutrino flux, nor terrestrial
modulation of the flux, has been observed, (d) a deficit of $^7$Be solar
neutrinos, if there is one, may still be explained by standard physics
and/or astrophysics (e) only future observations of spectral distortions
or flavor change of solar neutrinos in solar neutrino experiments, such
as Super-Kamiokande, SNO, Borexino and HELLAZ may establish that neutrino
properties beyond the minimal standard electroweak model are responsible
for the solar neutrino problem.

\section{Solar Neutrino Observations} 
\subsection{The Chlorine Detector}
The radiochemical chlorine detector (Davis 1966, Cleveland et al. 1998)
which contains 615 tons of
tetrachloroethylene, ${\rm C_2Cl_4}$, measures the production rate of
$^{37}$Ar by solar neutrinos through the reaction ${\rm\nu_e+^{37}Cl
\rightarrow ^{37}Ar+e^-}$ which has a threshold energy of 814 keV. The
detector is located at the Homestake gold mine at a depth of 1480 meters
underground ($\sim 4200$ meters water equivalent) in Lead, South Dakota,
USA. After an exposure time of 1 to 3 months the ${\rm^{37}Ar}$ atoms are
extracted from the target liquid and counted by observing their electron
capture decay back to ${\rm ^{37}Cl}$ (half life ${\rm T_{1/2}=35~d}$) in
a proportional counter. 
The ${\rm ^{37}Ar}$ production rate measured in 108 individual runs (runs
18 to 133) from 1970 to 1995 is plotted in Fig. 2. The mean capture rate of
solar neutrinos was 
\begin{equation}
{\rm <\sigma\phi_{\nu_e}>_{Cl} = 2.56\pm 0.16(stat)\pm 0.14(syst)~SNU}
\end{equation}
where 1 SNU=1 neutrino capture per second in $10^{36}$ target atoms. 

\subsection{The Light Water Cherenkov Detectors}
{\bf Kamiokande II}, the first imaging light water solar neutrino detector,
(later upgraded to Kamiokande III) was located in the Kamioka mine in Japan
(Fukuda et al. 1996).
It measured the Cherenkov light emitted by electron recoils produced by 
elastic scattering of solar neutrinos from electrons in the inner 
680 tons of a large tank filled with a total of 2180 tons of 
light water. For radioactive background reduction the threshold had to be 
set to a rather high electron recoil energy (7 MeV). Therefore, the 
detector was sensitive only to the upper end of the ${\rm ^8B}$ solar 
neutrino spectrum. During cumulative live time of 2079 days the detector 
yielded an average flux of 
\begin{equation}
{\rm \phi_{\nu_e}( ^8B)=2.80\pm 0.19(stat)
\pm 0.33(syst)~ \times 10^6~cm^{-2}s^{-1}},  
\end{equation}
as shown in Fig. 3. Within its experimental sensitivities KII+KIII  
has not detected temporal variation or spectral distortion of the
the ${\rm ^8B}$ spectrum.

\noindent
{\bf Super-Kamiokande} (SK) is a 500,00 tons imaging water Cherenkov 
detector whose inner 22,500 tons are used for the solar neutrino 
measurements (Fukuda et al 1998). The current energy  threshold is 6.5 
MeV. It started its 
operation on April 1 1996. Fig. 4 shows the distribution of events as 
function of the cosine of the angle of electrons recoiling from 
neutrino-electron scattering relative to the direction from the sun 
during live time of 504 days. The solid line is the best fitted
histogram due to the ${\rm ^8B}$ solar neutrino flux of (Suzuki et al. 1998)
\begin{equation} 
{\rm\phi_{\nu_e}( ^8B)= 2.44^{+0.06}_{-0.05}(stat)^{+0.09}_{-0.07}(syst)
~ \times 10^6~ cm^{-2}s^{-1}}
\end{equation}
and a constant background. The observed energy spectrum of the recoiling 
electrons is consistent with that expected from elastic scattering of 
${\rm^8B}$ solar neutrinos. In spite of the large number of events the 
statistics are not large enough yet to show the expected $\approx 6.5\%$ 
periodical variation in the flux due to the periodical variation 
in the distance of Earth from the sun, as can be seen from Fig. 5. 
The SK data also does not show any dependence on the path length of 
solar neutrino in Earth, as can be seen from Fig. 6. The SK limit on a 
Day-Night asymmetry is 
\begin{equation}
{\rm A={D-N \over D+N}=-0.023\pm 0.020(stat)\pm0.014 (syst)}.
\end{equation}

\subsection{The Gallium Detectors}  
The radiochemical gallium detectors measure the production rate of
$^{71}$Ge by solar neutrinos through the reaction ${\rm
\nu_e+ ^{71}Ga\rightarrow ^{71}Ge+e^-}$. The energy threshold, 232.2 keV,
is well below the maximum energy of the pp neutrinos, 420 keV.  After
exposure time of a couple of weeks the ${\rm^{71}Ge}$ atoms are 
extracted from
the target liquid and counted by detecting the  X-ray emission 
from their 
electron capture decay
back to ${\rm ^{71}Ga}$ (half life ${\rm T_{1/2}= 11.43~days }$) in a 
proportional counter. 

\noindent
{\bf The Gallex detector} (Hampel et al. 1996; Kirsten 1998)
is located in the Gran Sasso  underground laboratory in Italy. It contains
30.3 tons of gallium in ${\rm GaCl_3-HCl}$ solution. The neutrino produced
$^{71}$Ge atoms form the volatile compound ${\rm GeCl_4}$ which at the 
end of the exposure is swept out of the solution by a gas stream and 
converted into ${\rm GeH_4}$. The produced ${\rm ^{71}Ge}$ are counted by 
detecting their radioactive decay inside a proportional counter. 
The combined results of 65 individual GALLEX runs corresponding to data 
taking periods GALLEX I,II,III and IV as shown in Fig. 8
gave a production rate of
\begin{equation}
{\rm <\sigma\phi_{\nu_e}>_{Ga}=76.4\pm 6.3(stat)^{+4.5}_{-4.9}(syst)~SNU}.
\end{equation}
GALLEX has also conducted two ${\rm^{51}Cr}$ neutrino source experiments
to test the overwhole performance of the detector. The ratio between the 
measured ${\rm^{71}Ge}$ production rate due to the ${\rm^{51}Cr}$ source
and the expected rate from the known source strength was $1.00\pm 0.10$
and $0.83\pm0.10$, in the two experiments, respectively. 

\noindent
{\bf The Soviet American Gallium Experiment} (SAGE) is located in 
the Baksan Neutrino Observatory in the northern Caucasus of Russia at
a shielding depth of 4715 meter water equivalent. The detector uses 55 
tons of metallic gallium. After the metal is converted to solution the 
produced ${\rm ^{71}Ge}$ atoms are removed from the Gallium  and counted 
in a low background proportional counters by a procedure similar to that   
used by GALLEX. The combined results of 65 individual runs of SAGE
corresponding to data taking periods SAGE I,II,III,IV as shown in Fig. 
8 gave a production rate of (Abdurashitov et al 1996, Gavrin 1998) 
\begin{equation}
{\rm <\sigma\phi_{\nu_e}>_{Ga}=70\pm 6.3(stat)^{+4.5}_{-4.9}(syst)~SNU}.
\end{equation}
SAGE has also conducted a ${\rm ^{51}Cr}$ neutrino source experiment. The 
ratio between the measured ${\rm^{71}Ge}$ production rate due to the
${\rm^{51}Cr}$ source and the rate expected from the source strength was
$0.95\pm 0.11$. 

\section{The Luminosity Sum Rule}
Due to conservation of baryon number, electric charge, lepton flavor and
energy, the net reaction in the sun can be written as 
\begin{equation} 
2e^-+4p\rightarrow ^4{\rm He}+2\nu_e+26.732~MeV. 
\end{equation} 
Thus, the generation of $Q=26.732$ MeV in the sun is accompanied by the 
production of two $\nu_e$'s. If the
sun is approximately in a steady state with a nuclear energy 
production rate that equals its luminosity, then the total solar neutrino 
flux at Earth is (Dar and Nussinov 1991),
\begin{equation} 
\phi_{\nu_e}\approx
{2S_\odot\over Q-2\bar{E}_{\nu_e}}\approx 6.54\times 10^6~cm^{-2}~s^{-1}, 
\end{equation} 
where $S_\odot=L_\odot/4\pi
D_\odot\approx 1.367~kW~m^{-2}$ is the measured ``solar constant'' which
yields a solar luminosity $L_\odot\approx 4\pi D^2\approx 3.846\times
10^{33}~erg~s^{-1}$ for an average distance $D_\odot\approx 1.496\times
10^{13}~cm$ of Earth from the sun, and $\bar E_{\nu_e} = \Sigma
E_{\nu_e}(i)\phi_{\nu_e}(i)/\phi_{\nu_e}$ is the mean energy of solar 
neutrinos which
has been approximated by $\bar E_{\nu_e}(pp)\approx 0.214$ MeV, the mean 
energy of
the pp solar neutrinos that dominate the solar neutrino flux. Eq. (2) can 
also be rewritten as a sum rule,
\begin{equation} 
\Sigma_i (Q/2-\bar E_{\nu_e}(i)) \phi_{\nu_e(i)}\approx S_\odot. 
\end{equation}
The summation extends over all the neutrino producing reactions with 
$\bar E_{\nu_e}=0.265$, 1.442, 0.814, 6.710, 0.707, and 0.997 MeV for
the pp, pep, $^7$Be, $^8$B, $^{13}$N and $^{15}$O neutrinos, respectively.
If the small pep flux, which is proportional to the pp flux,  is included 
in the pp flux, and if the very small $^8$B neutrino flux is neglected 
then the solar luminosity sum rule can be rewritten as       
\begin{equation}
{\rm 0.9800\phi_{\nu_e}(pp)+0.939\phi_{\nu_e}(Be)+0.936\phi_{\nu_e}(NO)
\approx 6.377},  
\end{equation} 
where ${\rm \phi_{\nu_e}(i)}$ are in units of ${\rm 10^{10}~cm^{-2}s^{-1}}$.

\section{Helioseismological Constraints} 
Accurate ground based (e.g., Hill
et al. 1996 and references therein) and space based measurements aboard
SOHO (e.g., Turck-Chieze et al. 1997 and references therein) of solar
photospheric oscillation frequencies provided detailed information about
the structure of the solar interior (e.g., Christensen-Dalsgaard 1996). In
particular the base of the convection zone has been determined to be at
${\rm R_{cz}\approx (0.713\pm 0.003)R_\odot}$ (Basu and Antia 1997) and
the photospheric helium abundance has been inferred to be ${\rm
Y_s=0.249\pm 0.003}$ (Basu and Antia 1995). Helioseismology is generally
in good agreement with the standard solar models (see e.g. 
Christensen-Dalsgaard 1996). However, there are systematic deviations
between the helioseismology determination of the sound speed in the sun
and that predicted by the SSM (see, e.g., Fig. 10.) which are similar 
to all SSMs and whose origin is not clear. Moreover, helioseismology is
sensitive only to the average local properties in the sun (temperature,
density, average molecular weight) but not to the rates of the rare
nuclear reactions in the sun which produce the pep, $^7$Be, $^8$B and NO
solar neutrinos. Therefore, helioseismology should not be used to
argue that the SSM predict correctly these solar neutrino fluxes.

\section{Model Independent considerations} 
Counting rates in solar neutrino experiments are formally given by 
\begin{equation}
R=N_a\Sigma_i \phi_{\nu_e}(i)\int_{E_0}(dn_{\nu_i}/dE)\sigma_{\nu a}
(E)dE 
\end{equation}
where $N_a$ is the number of ``active'' atoms in the detector,
$\sigma_{\nu a}(E)$ is their cross section for neutrinos with energy E,
$E_0$ is the threshold energy of the detector, 
$dn_{\nu_i}/dE$ is the  energy spectrum and $\phi_{\nu_e}(i)$ is the total 
flux of neutrinos from reaction $i$ in the sun. 
Both, $dn_{\nu_i}/dE$ and $\sigma_{\nu a}$ follow from
the standard electroweak theory and are essentially independent of the sun: 
$dn_{\nu_i}/dE$ is practically the standard $\beta$-decay
spectrum for the $\beta$-decays 2p$\rightarrow De^+\nu_e$,
$^8$B$\rightarrow 2\alpha e^+\nu_e$, $^{13}$N$\rightarrow
^{13}$C$e^+\nu_e$ and $^{15}$O$\rightarrow^{15}$N$e^+\nu_e$ and is a 
$\delta$-function for the electron captures 
$e^-$$^7$Be$\rightarrow\nu_e~^7$Li
and $pep\rightarrow D\nu_e$.) Thus conclusive evidence
for new electroweak physics can be provided only by detecting
at least one of the following signals:
(1) Spectral distortions of the $\beta$-decay spectra of solar neutrinos.
(2) Solar neutrino flavors other than $\nu_e$.         
(3) Terrestrial modulations of the solar neutrino flux. 
(4) A clear violation of the luminosity sum rule.

\noindent
So far, no such  conclusive evidence has been provided by the solar
neutrino experiments. 

\noindent
{\bf 1. Spectral Distortion.} At present only Super-Kamiokande can
measure the spectrum of solar neutrinos (above 6.5 MeV). Within their
statistics and systematic uncertainties the energy distribution of the
detected electrons which are scattered by solar neutrinos is consistent
with that expected from an undistorted spectrum of $^8$B neutrinos. This
can be seen from Fig. 7. A ``hint'' for a spectral distortion may exist
in the SK data, but it depends strongly on events beyond the kinematical
limit which are attributed to the detector energy resolution. Also the
neutrino spectrum near the ``end point'' is not known very well because
the $^8$B decays into a virtual short lived $2\alpha$ state which has 
a large energy spread. 
 
Super-Kamiokande, which has been running since April 1, 1996, will finally
have a much larger statistics and perhaps a lower threshold energy. These,
perhaps, will be able to provide more conclusive evidence. 

\noindent
{\bf 2. Neutrino Flavor Change.} The radiochemical experiments are blind to
neutrino flavors other than that of $\nu_e$'s. SK is sensitive also
to $\nu_\mu$'s and $\nu_\tau$'s but it cannot distinguish between the 
neutrino flavors at solar energies. Only future experiments, such as SNO,
will be able to obtain information on the flavor content of the
solar neutrino flux.     

\noindent
{\bf 3. Terrestrial Modulations}.
In spite of the large number of events collected by SK the statistics are
not large enough yet to show the expected $\approx 6.5\%$ periodical
variation in the flux due to eccentricity of the Earth's orbit around the
sun, as can be seen from Fig. 5.  The SK data also show no dependence on
the path length of solar neutrino in Earth, as can be seen from Fig. 6, no
day-night effect and no winter-summer difference. In particular, its limit
on the Day-Night asymmetry is ${\rm A=(D-N)/(D+N)=-0.023\pm 0.025}$. 

\noindent
{\bf 4. Violation of the Luminosity Sum Rule}.
A clear violation of the solar luminosity sum rule could prove that lepton
flavor is not conserved. The ``minimal'' expected signal in gallium which
follows from the luminosity sum rule is obtained by assuming that 
all the solar neutrinos are pp neutrinos. If the 
mean cross section for the capture of the pp neutrinos in gallium is
$\sigma\approx (1.17\pm 0.03)\times 10^{45}~ cm^{-2}$, it yields 
a minimal signal of $76\pm 2$ SNU in $^{71}$Ga. In fact, the $^8$B
solar neutrino flux, which is observed by SK  contributes additional
${\rm <\sigma \phi(^8B)>_{Ga}=5.85\pm 1.75}$ SNU to the $^{71}$Ge 
production rate and the minimal expected signal in $^{71}$Ga is
$82\pm 3$ SNU.  It is consistent, within the experimental and
theoretical uncertainties, with the $76.4\pm 8$ SNU production 
rate of ${\rm ^{71}Ge}$ by solar neutrinos in gallium which was measured by
GALLEX. The gallium experiments, however, appear to leave no extra room
for the contribution from ${\rm ^7Be}$ solar neutrinos. 

A further indication that the $^7$Be and perhaps also the CNO neutrinos
are missing is provided by the Cl experiment. The expected 
production rate of $^{37}$Ar in $^{37}$Cl by the $^8$B solar neutrino 
flux that was measured by Super-Kamiokande is ${\rm <\sigma 
\phi_{\nu_e}>_{Cl}=2.68\pm 0.15~ 
SNU}$. It is consistent with ${\rm <\sigma \phi_{\nu_e}>_{Cl}=2.56\pm 0.21~ 
SNU}$, the total $^{37}$Ar production  rate in $^{37}$Cl, 
as measured at Homestake but leaves no room for the contribution
from the $^7$Be solar neutrino flux.

The SK measurement (3) and the solar luminosity sum rule (10)
two observational constraints on the solar neutrino fluxes.
Two additional constraints are provided by the gallium and chlorine
experiments. Assuming that the neutrino capture cross sections
are well represented by their theoretical estimates (e.g., Bahcall
1989; 1998 and references therein) one can write them approximately as 
\begin{eqnarray} 
{\rm Ga} &:~&{\rm (11.7\pm 0.3)\phi_{\nu_e}(pp)+ (71.7\pm 
5.0)\phi_{\nu_e}(^7Be)+(2.40\pm 0.78)\times 10^4 \phi_{\nu_e}(^8B)}
 \nonumber \\
&~+& {\rm (87\pm 12) \phi_{\nu_e}(NO)=77.5\pm 7.8} 
\end{eqnarray}
\begin{eqnarray}
{\rm Cl}&:~&~~{\rm  (1.11\pm 0.04)\times 
10^4\phi_{\nu_e}(^8B)+(2.4\pm)\phi_{\nu_e}(^7Be)} \nonumber\\
&~+&{\rm (16\pm)\phi_{\nu_e}(pep)+ (4.2\pm)\phi_{\nu_e}(NO)=2.56\pm 0.25}
\end{eqnarray}

If the theoretical estimates of the cross sections for nuclear
capture of solar neutrinos represent well their true values
then the only physical solution of the four constraints is
\begin{eqnarray} 
& &{\rm \phi_{\nu_e}(pp+pep)\approx 6.5 \times 10^{10} cm^{-2}s^{-1},~
\phi_{\nu_e}(^8B)\approx 2.44\pm 0.11 \times 10^6 cm^{-2}s^{-1}},
\nonumber\\
& &{\rm ~~\phi_{\nu_e}(^7Be)\ll\phi_{\nu_e}^{SSM}(^7Be),~~~~~~~~~
\phi_{\nu_e}(NO)\ll\phi_{\nu_e}^{SSM}(NO)}.
\end{eqnarray}
The confidence level of this solution cannot be quantified  
in a reliable way because of the unknown origin and nature of all
the systematic errors in both the theoretical cross sections and the
experimental results. The only reliable conclusion is that 
${\rm \phi_{\nu_e}(^7Be)}$ and ${\rm\phi_{\nu_e}(NO)}$ appear to be 
strongly suppressed compared with their standard solar model estimates.

\section{Is There a $^8$B Solar Neutrino Problem?} Table I presents a
comparison between the solar neutrino observations and the SSM predictions
of Bahcall and Pinsonneault 1995 (BP95), Bahcall, Basu and Pinsonneault 1998
(BSP98), Brun, Turck-Chieze and Morel 1998 (BTM98)
and Dar and Shaviv 1996 (DS96). Although BP95 and BSP98 predict a
$^8$B solar neutrino flux that is approximately 2.2 and 2.7 times,
respectively, larger than that observed by SK, DS96 predict a flux
consistent with that observed by SK. The differences between BP95, BSP98
and DS96 are summarized in Table II (for details see Dar and Shaviv 1996).
The difference between the predicted $^8$B flux are mainly due to the use
of different nuclear reaction rates in DS96, differences in the calculated
effects of diffusion and differences in the initial solar chemical
composition assumed in the two calculations. They reduce the predicted
$^8$B flux relative to those in BP95 (BSP98) by approximate factors of 0.55
(0.70), 0.81, and 0.95, respectively. The remaining differences are
mainly due to inclusion of partial ionization effects, premain sequence
evolution and deviations from complete nuclear equilibrium in DS96 which
were neglected in BP95 and BSP98, and due to different numerical methods,
fine zoning and time steps used in the two calculations:

\subsection{Initial Chemical Composition}
The initial chemical composition of the sun influences 
significantly the solar evolution and consequently the present density, 
chemical
composition and temperature in the solar core that determine the solar
neutrino fluxes. In particular, the radiative opacities, which determine
the temperature gradient in the solar interior, are very sensitive to the
abundance of heavy elements which are not completely ionized in the sun. 
Since the initial chemical composition of the sun is unknown, one must
infer it indirectly, e.g., from the chemical composition of the solar
photosphere, the primitive early solar meteorites and the local
interstellar medium (ISM). 

The solar photospheric abundances have changed only slightly during the
solar evolution by gravitational settling, diffusion and turbulent mixing
in the convective layer and by cosmic ray interactions with the surface
material. Therefore, in principle, one can adjust the initial
chemical composition of the sun in the SSM to yield its measured photospheric
composition. Unfortunately, the photospheric abundances of most elements
are still not known to sufficient accuracy and there is no direct
spectroscopic information on the photospheric abundance of $^4$He.
Consequently, the initial mass fraction ${\rm Y_i}$ of $^4$He in the sun
has been treated in the SSM as an adjustable parameter. Recently, however,
the photospheric mass fraction of $^4$He has been inferred from
helioseismological measurements of the sound speed in the convective solar
layer.  The best estimated value is now ${\rm Y_s}=0.249\pm 0.003$
(Basu and Antia 1997). Since the photospheric mass ratio of metals 
((${\rm A>4}$) to hydrogen is (Grevesse,
Noels and Sauval 1996) ${\rm Z/X}=0.0244$, one obtains that ${\rm
X_s=0.733}$ and ${\rm Z_s}= 0.0178$. It is, however, important to note
that the metallicity in the present local ISM, essentially measured from
analyses of the Orion nebula and of nearby B stars (Gies and
Lambert 1992; Cunha and Lambert 1992; Wilson and Rood 1994; Mathis 1996)
is lower than the value obtained from the photospheric abundances. This is
in contradiction with galactic chemical evolution models which predict an
increase with time of metallicity in the ISM. Moreover, gravitational
settling and diffusion decrease the solar surface  metallicity with 
time.

The meteoritic elemental abundances are known with much better accuracy.
Beside the noble gases, and H, C, N and O which were able to form highly
volatile molecules or compounds and escape condensation, all the other
elements are believed to have condensated completely in primitive early
solar system meteorites. Therefore, their initial relative abundances in
the sun are expected to be well represented by their values in type I
carbonaceous chondrites. If diffusion and gravitational settling have not
changed their ratios significantly, then the relative abundances of these
elements in meteorites must be similar to their photospheric values. Over
the past decades there have been many initial disagreements between the
meteoritic and photospheric abundances. In nearly all cases, when the
atomic data were steadily improved and the more precise measurements were
made, the photospheric values approached the meteoritic values. The
photospheric abundances are now as a rule in very good agreement with the
meteoritic values if the conversion factor from the solar abundance scale 
${\rm N_H=10^{12}}$ to the meteoritic scale, ${\rm N_{S_i}=10^6}$ 
is ${\rm R=log(sol)/met)=1.56 }$, as can be seen from Table III borrowed 
from Grevesse, Noels and Sauval 1996. 

In our SSM, ${\rm Y_i}$ was
left as an adjustable parameter.  The initial solar heavy metal abundances
were assumed to be equal to the meteoritic (CI chondrites) values of
Grevesse, Noels and Sauval (1996). The overwhole initial metallicity ratio
${\rm Z_i/X_i}$ and the initial H, C, N, O and Ne abundances were adjusted
to yield their present photospheric values (${\rm Z_s/X_s=0.0244}$ and the
values quoted in Table III, respectively).  Our SSM yields ${\rm
Y_s=0.238\pm 0.05}$ in agreement with the helioseismological estimates.

\noindent 
The photospheric abundances of $^7$Li, however is smaller by
a factor of nearly 140 than its meteoritic
abundance. The origin of such large difference is still not clear. It
cannot be explained by nuclear burning during the Hayashi phase although
significant lithium burning does takes place during this phase. It may
be explained by rotational mixing (e.g., Richard et al 1996). Although the
initial solar (meteoritic) abundances of lithium is
very small and do not play any significant role in solar evolution, its
depletion perhaps can provide a clue to the real history of the convection
zone and the sun.

\subsection{Nuclear Reaction Rates} 
The nuclear reaction rates for most stellar reactions are inferred by
extrapolating measurements at higher energies to stellar reaction 
energies. The cross sections at center of mass energies well below 
the Coulomb barrier are usually parametrized as 
\begin{equation} 
\sigma(E)={S(E)\over E}e^{-2\pi\eta(E)} 
\end{equation}
where $\eta=Z_1Z_2e^2/\hbar v$ is the Sommerfeld parameter, $v$ is the 
relative velocity of the colliding nuclei in the initial state,
$Z_1$ and $Z_2$ are their charge numbers and $E$ is their center of 
mass energy. The exponent is an approximate (WKB) form  for the 
penetration probability of the Coulomb barrier in the initial state.   
The ``astrophysical S factor'' is expected to vary only slowly with energy.
It is usually extracted either from a polynomial fit to experimental data
at low energies or from theoretical calculations normalized
to the experimental data.

The uncertainties in the nuclear reaction rates at solar conditions are
still large due to (1) uncertainties in the measured cross sections at
laboratory energies, (2) uncertainties in their extrapolations to lower
solar energies, (3) uncertainties in dense plasma effects (screening,
correlations, fluctuations and deviations from pure equilibrium
distributions) on reaction rates.  Rather than averaging measured cross
sections that differ by many standard deviations, DS96 used for their
extrapolations only the most recent and consistent measurements of the
relevant nuclear cross sections. For sub-Coulomb reactions that take place
when the colliding nuclei are far apart, the Optical Model and the
Distorted Wave Born Approximation give a reliable description of their
energy dependence and can be used for extrapolating the measured
sub-Coulomb cross sections to solar energies. 

The ``astrophysical S factors'' which were used in BP95, BSP98 and DS96  
are compared in Table II. The origin of the differences are as follows:

\noindent {\bf ${\rm S_{11}(0)}$}.
The value advocated by Adelberger et al. (1998) and used in BSP98 is based
essentially on the updated calculation of ${\rm S_{11}(0)}$ by Kamionkowski
and Bahcall (1994). The authors considered many small corrections.
However, they ignored the {\it screening effect of the solar plasma
electrons in the outgoing channel}. Screening corrections in the nuclear
reaction rates were included in BP95 and BSP98, and in all the published
SSM calculations, only in the incoming channel. That is justified for
radiative captures like ${\rm ^3He+ ^4He\rightarrow ^7Be+\gamma}$ and
${\rm p+ ^7Be\rightarrow ^8B+\gamma}$ because the photon is chargeless. The
emitted positron in the``beta decay'' ${\rm pp\rightarrow D+e^++\nu_e}$ of
the two fusing protons sees essentially a Deuterium nucleus screened by
the plasma electrons in the Debye sphere left by the two protons (the mean
velocity of the emitted positron is much larger than the mean velocity of
the electrons in the Debye sphere). If the wave function of the ejected
positron in the Coulomb field of the Deuteron is calculated from the Dirac
equation with the Debye screening potential around two fusing protons, one
obtains that their fusion rate in the sun is enhanced by approximately
1.75\% (Dar and N. Shaviv, unpublished). Consequently, the ``bare'' value
of ${\rm S_{11}(0)}$ in SSM calculations that do not include electron
screening in the outgoing channel must be increased by 1.75\% as in DS96. 

\noindent 
{\bf ${\rm S_{17}(0)}$}. Recent low energy measurement of the cross 
section for
the reaction ${\rm p+^7Be\rightarrow ^8B+\gamma}$ by Hammache et al. (1998)
are consistent with the measurements of Vaughn et al. (1970) and of
Filippone et al. (1983a,b) which reached energies as low as 134 keV. These
measurements disagree with the older measurements of Kavanagh (1960), Parker
(1966, 1968) and Kavanagh et al. (1969). 

Because of the low binding energy, the radiative capture of p by $^7$Be
takes place well outside the range of the nuclear forces. Therefore, at
low energies the dependence of the cross section on energy is well
described by the optical model. When applied to the experimental data of
Vaughn (1970); Filiponne (1983a,b) and Hammache et al. (1998) it yields
${\rm S_{11}(0)=17.8\pm 1.0~eV\cdot b}$ (see also Barker 1995; Nunes et al
1997). This value is consistent with the indirect measurements, ${\rm
S_{11}(0)=16.7\pm 3.2~eV\cdot b}$ through Coulomb dissociation of ${\rm
^8B}$ (Motobayashi et al 1994) and ${\rm S_{11}(0)=17.6\pm 1~eV\cdot b}$
through proton transfer reactions (Xu et al. 1994). The mean value ${\rm
S_{17}(0)=17.5\pm 1.0 ~eV\cdot b}$ is consistent with the value ${\rm
S_{17}(0)=17~eV\cdot b}$ used in DS96, but it is smaller than both ${\rm
S_{17}(0)=22.4~eV\cdot b}$ used in BP95 and ${\rm S_{17}(0)=19~eV\cdot b}$
that is advocated by Adelberger et al. 1998 and used in BSP98. 

\noindent
{\bf ${\rm S_{33}(0)}$}. The cross section for the reaction ${\rm
^3He+^3He \rightarrow ^4He+2p}$ has recently been measured 
at the Laboratory for Underground Nuclear Astrophysics (LUNA) at 
energies covering the Gamow peak around ${\rm E_{G}=21.9~keV}$
(Junker et al. 1998). At such small
lab energies screening by atomic electrons enhances considerably the
bare nuclear cross section. In the Born-Oppenheimer adiabatic
approximation, $U_e$, the gain in kinetic energy by the colliding 
nuclei, is bounded by the change in the total binding energy of the atomic
electrons when they occupy the atomic ground state of the
${\rm^3He+^3He}$ ``nuclear molecule''. For ${\rm ^3He^{++}}$ ions 
incident on $^3$He gas target, $U_e\approx$ 240 eV.  From
eq. (15) one obtains that the screening enhancement of the cross section
for $U_e\ll E$ is given approximately by 
\begin{equation}
\sigma=\sigma_{exp} e^{\pi\eta U_e/E}. 
\end{equation} 

We stress that one must use a consistent treatment of electron screening
enhancement of nuclear cross sections in the lab and in the solar plasma.
In particular, $U_e$ cannot be taken as an adjustable parameter when
fitting ${\rm S(E)}$ to the lab measurements but must be fixed to its
adiabatic value $U_e=240$ eV in order to be consistent with the ``weak
screening'' prescription that is used in the SSM. For $U_e=240$ eV, a best
fitted ${\rm S(E)}$ between 20.7 keV and 1080 keV to the cross section
measurements that are consistent at overlapping energies (Dwarkanath and
Winkler 1971, Krauss et al. 1987, Greife et al.  1994 and Junker et al.
1998) yields the values ${\rm S_{33}(0)=5.60~MeV\cdot b}$, ${\rm
S_{33}'(0)=-4.1 b}$ and ${\rm S_{33}''(0)=4.60~MeV^{-1}\cdot b}$.  The fit
is shown in Fig. 9. Note that if one extracts S(E) directly from the
LUNA data at the Gamow peak (Junker et al. 1998) one obtains ${\rm
S_{33}(E_{G})=5.75~MeV\cdot b}$. The value ${\rm S_{33}(0)=4.99~MeV\cdot
b}$ was used in BP95, the value ${\rm S_{33}(E_{G)}=5.30~MeV\cdot b}$ that
was recommended by Adelberger et al. 1998 was used in BSP98, while the
value ${\rm S_{33}(0)=5.6~MeV\cdot b}$ was used in DS96. 

\noindent {\bf ${\rm S_{34}(0)}$}.  There are six published measurements
of the cross section for the reaction ${\rm ^3He+^4He\rightarrow
^7Be+\gamma}$ that are based on the detection of the prompt $\gamma$-rays
(Parker and Kavanagh 1963; Nagatani et al. 1969; Kr\"awinkel et al. 1982;
Osborne et al. 1982,1984; Alexander et al. 1984; Hilgemeier et al. 1988)
and three that are based on the late decay of $^7$Be (Osborne et al.
1982,1984; Robertson et al. 1983; Volk et al. 1983). There is a systematic
discrepancy of more than $3\sigma$ between these two data sets whose
origin is not clear. When theoretical models are used to extrapolate the
direct measurement to low energies they yield a weighted mean ${\rm
S_{34}(0)=0.507\pm 0.016~ MeV\cdot b}$, while the activation measurements
yield ${\rm S_{34}(0)=0.579\pm 0.024~MeV\cdot b }$. The weighted mean of
all experiments, ${\rm S_{34}(0)=0.53\pm 0.05}$ (Adelberger et al. 1998)
was used in BSP98.  However, the radiative capture ${\rm
^3He+^4He\rightarrow^7Be+\gamma}$ does not take place well outside the
range of the nuclear forces ($Z_1Z_2e^2/E_{B}\approx 3.6~fm<{\rm
R(^3He)+R(^4He)}$). Thus, the ability of the theoretical models to 
predict correctly the low energy dependence of ${\rm S_{34}(E)}$ is 
questionable. 
Moreover, the energy dependence of the cross section which they predict
disagrees with that observed in the measurements of Kr\"awinkel et al. 1982
who used gas targets, have low statistical errors and extend to the
lowest energies. If one uses a polynomial fit of ${\rm S_{34}(E)}$ to
these data, and than uses it to extrapolate the other measurements to low 
energies, one obtains from the direct measurements a weighted mean ${\rm
S_{34}(0)=0.45~MeV\cdot b}$ which was used in DS96.

\subsection{Diffusion} Diffusion, caused by density, temperature,
pressure, chemical composition and gravitational potential gradients plays
an important role in the sun since it modifies the local chemical
composition in the sun. The agreement between helioseismology and the SSM
is improved when diffusion is included (see e.g., Fig. 10), but inclusion
increases the discrepancy between the SSM and the solar neutrino
observations.  The relative changes in SSM predictions due to diffusion of
all elements are summarized in Table IV. While BP95 and BSP98 found a
rather large increase in the predicted $^7$Be, $^8$B, $^{13}$N, $^{15}$O
and $^{17}$F solar neutrino fluxes; 14\%, 36\%, 52\%, 58\%, and 61\% which
result in 36\%, 33\%, 9\% increases in their predicted rates in
Super-Kamiokande, Homestake, and in GALLEX and SAGE, respectively, DS96
found only a moderate increase due to diffusion, 4\%, 10\%, 23\%, 24\% and
25\%, respectively, in the above fluxes, which result in 10\%, 10\% and
2\% increase in the predicted rates in Kamiokande, Homestake, and in
GALLEX and SAGE, respectively. Although the two diffusion calculations
assumed a different initial solar chemical composition (see below) and BP95
approximated the diffusion of all elements heavier than $^4$He by that of
fully ionized iron (the DS calculations followed the diffusion of each
element separately and uses diffusion coefficients calculated for the
local ionization state of each element in the sun as
obtained from solving the local Saha equations), these cannot fully
explain the above large differences.  Independent diffusion calculations
by Richard et al. (1996) obtained similar results to those obtained in
DS96 as can be seen from Table IV (we interpolated the results from the
two models of Richard et al. (1996) to the initial chemical composition
assumed in DS96). Note that internal magnetic fields can suppress
diffusion significantly.

\section{New Neutrino Physics?} Standard solar models, like DS96, perhaps
can explain the results reported by Kamiokande and Super-Kamiokande. 
However, if the neutrino absorption cross sections assumed by the
radiochemical experiments are correct, then present standard solar models
cannot explain the absence of the expected contributions of the $^7$Be and
CNO solar neutrinos to the $^{37}$Ar production rate in $^{37}$Cl and to
the $^{71}$Ge production in $^{71}$Ga. Consequently, many authors have
claimed that the the solar neutrino observations imply neutrino properties
beyond the minimal standard electroweak model (e.g., Bahcall and Bethe
1991).

\noindent {\bf Neutrino Magnetic Moments.} Some authors have suggested
that neutrinos may have anomalous magnetic moments large enough so that
the solar internal magnetic field in the sun can flip the neutrino helicity
and convert part of the left handed weakly interacting solar neutrinos into
noninteracting right handed neutrinos (e.g., Okun et al. 1986, Lim and
Marciano 1988, Akhmedov 1988). However, the high statistics solar neutrino
measurements of SK do not show the time variation of the solar neutrino
flux which is predicted by a magnetic helicity flip interpretation of the
solar neutrino anomaly. 

\noindent 
{\bf Neutrino Oscillations}: 
Mikheyev and Smirnov (1985) have discovered that
neutrino oscillations in matter (Wolfenstein 1978, 1979) can 
lead to resonant conversion of neutrino flavor (the MSW effect) in 
the sun and explain the
solar neutrino observations quite neatly. It requires only a
natural extension of the minimal standard electroweak theory and it is
based on a simple quantum mechanical effect. Many authors have carried out
extensive calculations to determine the neutrino mixing parameters which
can bridge between the predictions of the standard solar models and the
solar neutrino observations. 
The neutrino mixing parameters can also be deduced analytically
directly from the solar neutrino observations (e.g., Dar and Nussinov 1991;  
Dar 1993): If
the $\nu_e$ is mixed with the $\nu_\mu$ (or the $\nu_\tau)$ with a vacuum
mixing angle $\theta\ll1$ and a mass difference $\Delta
m^2=m_{\nu_\mu}^2-m_{\nu_e}^2$, then solar $\nu_e$'s which are produced in
the sun can flip their flavor on their way out of the sun if they
encounter electron density
\begin{equation}
n_e={\Delta m^2 c^4 cos 2\theta\over 2\sqrt{2}G_FE_{\nu_e}}\approx
    {\Delta m^2 c^4 \over 2\sqrt{2}G_FE_{\nu_e}}.
\end{equation} 
The  probability for the resonant flavor flip (the MSW effect)
is given approximately by (e.g., Haxton 1986; Parke 1986; Dar et al. 1987) 
$P(\nu_e\rightarrow \nu_\mu \approx 1-e^{-\epsilon/E_\nu}$ where
\begin{equation}
\epsilon\approx 
{\pi H\Delta m^2 c^4 sin^2 2\theta \over 4\hbar c\cdot cos2\theta}\approx 
{\pi H\Delta m^2 c^4\theta^2 \over \hbar c}, 
\end{equation}
with $H=-n_e/(dn_e/dr)$ being the density scale-height at the resonance.

Strong suppression of the contribution from the pep, $^7$Be and CNO
solar neutrinos to the Cl experiment requires a complete flavor change of
these neutrinos in the sun, i.e., that they encounter a resonant density
with $\epsilon\gg E_{\nu_e}\sim 1~$ MeV. But the results from GALLEX and
SAGE suggest that the lower energy pp solar neutrinos evade such a flavor
flip. This is possible if the resonance condition (17) is satisfied
for the pep, $^7$Be and CNO solar neutrinos ($E_{\nu_e}\geq 0.862$ MeV)
but not for the pp neutrinos ($E_{\nu_e}\leq 0.420$ MeV). Since the
maximal (central) electron density in the sun is $n_e\approx 10^2{\rm N_A}$ 
the last condition reads (e.g., Dar 1993)
\begin{equation}
 0.5\times 10^{-5}{\rm eV^2}\leq \Delta m^2c^4\leq 1 \times 10^{-5}
{\rm eV^2}
\end{equation}
For a given $\Delta m^2$, eq. (18) can be used to adjust the mixing angle   
to produce the required suppression of the SSM $^8$B solar $\nu_e$ flux:
\begin{equation} 
\Delta m^2 sin^2 2\theta\approx 3\times 10^{-8}{\rm eV^2c^{-4}},\\
\Delta m^2 sin^2 2\theta\approx 1\times 10^{-8}{\rm eV^2c^{-4}},
\end{equation}
for respectively, the BSP98 and the DS96 standard solar models.
  
\section{Are The $^7$Be Solar Neutrinos Missing ?} 
Electron capture by $^7$Be into the ground state of $^7$Li produces 862
keV neutrinos. The threshold energy for neutrino absorption by $^{37}$Cl
is 814 keV. Thus, absorption of $^7$Be neutrinos by $^{37}$Cl produces 48
keV electrons.  The maximum energy of the pp solar neutrinos is 420 keV.
The threshold energy for neutrino absorption in $^{71}$Ga $(3/2^-)$ is 233
keV into the ground state $(1/2^-)$ and 408 into its first excited state
$(5/2^-)$. The produced electrons have therefore energies below 187
and 12 $keV$, respectively. If the theoretical cross sections for neutrino
absorption near threshold overestimate their true values significantly 
then the predicted rates will significantly overestimate the expected
signals in the Chlorine and Gallium experiments.
\noindent
An indication that final state interactions effects are not completely
understood is provided by Tritium $\beta$-decay.  Although final state
interactions in Tritium $\beta$-decay have been studied extensively,
they do not explain well the end-point $\beta$-decay spectrum ($E_e\sim
18.6~keV$). In all recent measurements, the measured spectrum yields a
negative value for the fitted squared mass of the electron neutrino. 
Final state interactions effects (screening of the nuclear charge by
atomic electrons, exchange effects, radiative corrections, nuclear recoil
against the electronic cloud, etc) in neutrino captures near threshold in
$^{37}$Cl and $^{71}$Ga may be much larger because their Z values are much
larger and because the de Broglie wave lengths of the produced electrons
are comparable to the Bohr radii of the atomic K shells in Cl and Ga. If
final state interactions reduce considerably the near threshold absorption
cross sections of pp neutrinos in $^{71}$Ga (making room for the expected
contribution of $^7$Be solar neutrinos in Gallium) and of $^7$Be
neutrinos in $^{37}$Cl, perhaps they can make the solar neutrino
observations of Kamiokande and the Homestake experiment compatible.
Such an explanation of the solar neutrino problem implies that experiments
such as BOREXINO and HELLAZ will observe the full $^7$Be solar neutrino
flux.

\section{Astrophysical Solutions To The SNP}
\bigskip
Even if the $^7$Be solar neutrino flux is strongly suppressed, it does not
eliminate standard physics solutions to the solar neutrino problem: 
\noindent
The ratio between the fluxes of $^7$Be and $^8$B solar neutrinos 
is given by  
\begin{equation} 
R={\phi_{\nu_\odot}(^7{\rm Be})\over \phi_{\nu_\odot}(^8{\rm B})}
 = {\int n_e n_7<\sigma v>_{e7}d^3r\over
   \int n_p n_7<\sigma v>_{p7}d^3r}.
\end{equation}
Because of the decreasing temperature and Be7 abundance as function of
distance from the center of the sun on the one hand, and the $\sim r^2$ 
increase in radial mass on
the other, the production of $^7$Be and $^8$B solar neutrinos
in the SSM peaks around an effective radius, $r_{eff}\approx
0.064R_\odot$ ($r_{eff}$ is approximately the radius within which 50\% of
the flux is produced) . The SSM also predicts a ratio of electron to
proton densities near the center of the sun, $n_e/n_p\sim 2$, consistent
with helioseismology observations.  Consequently, the SSMs predict  
\begin{equation}
R\approx {2<\sigma v>_{e7}\over <\sigma v>_{p7}} \approx 1.27\times 
10^{-14}S_{17}^{-1}F_{17}^{-1}T_7^{1/6}e^{47.625/T_7^{1/3}},
\end{equation}      
where $F_{17}$ is the screening correction to the p-capture rate by $^7$Be,  
$T_7$ is the temperature in $10^7K$ at the effective radius and   
$S_{17}$ is in $eV~barn$ units.
The SSMs yield $T_7(r_{eff})\approx 1.45$. Using $S_{17}(0)=17~eV~b$  
and  $\phi_\odot(^8{\rm B})=2.44\times 10^6~cm^{-2}~s^{-1}$ as observed
by Super-Kamiokande, one can reproduce the SSM prediction (e.g., Dar and 
Shaviv 1996) 
\begin{equation}
\phi_{\nu_\odot}(^7{\rm Be})=R 
\phi_{\nu_\odot}(^8{\rm B})\approx 3.7\times 10^9~cm^{-2}~s^{-1}. 
\end{equation}
Astrophysical solutions of the solar neutrino problem aim towards
suppressing the value of R. Three alternatives are currently investigated:
 
\noindent {\bf Plasma Physics Effects}: The effects of the surrounding
plasma on nuclear reaction rates in dense stellar plasmas, and in
particular on proton and electron capture by $^7$Be in the sun are known
only approximately. In order to explain the deficit of $^7$Be solar
neutrinos, without much affecting the SSM, plasma screening effects must
reduce considerably the electron/proton capture ratio by $^7$Be, relative to
the predictions of the weak screening theory (Salpeter and Van Horne
1969). The screening enhancement of bare nuclear cross sections is not
well understood even in laboratory measurements with gas targets. Also the
applicability of the weak screening theory to the dense plasma in the
solar core is questionable.  Moreover, correlations and fluctuations,
which are neglected in the weak screening theory can affect strongly the
screening enhancement of nuclear reaction rates in the solar core. This
possibility is currently studied, e.g., by Shaviv and Shaviv (1998) using
numerical methods. Because of accidental cancellations the weak screening
corrections to the rates of all nuclear reactions do not change the
predicted $^8$B solar neutrino flux, but perhaps a more exact treatment of
screening may change R considerably. 

\noindent
Because the sub-Coulomb nuclear reactions in the core of the sun take
place mainly between nuclei with kinetic energies much larger than their
mean kinetic energies, their rates are very sensitive to the high energy
tail of their velocity distribution in the sun.  Diffusion, radiative
flows, energetic nuclear products, internal fluctating (equipartition ?)
electric and magnetic fields and other collective effects may change the
assumed Maxwell-Boltzmann tails of the energy distribution of the
energetic particles in the core of the sun. This may shift the position of
the Gamow peaks for the nuclear reaction rates and change considerably the
ratios between nuclear reaction rates in the sun which have very different
temperature dependence (see, e.g., Kaniadakis et al. 1998). 

\noindent 
In principle, collective plasma physics effects, such as very strong
magnetic or electric fields near the center of the sun, may also polarize the
plasma electrons, and affect the branching ratios of electron capture by
$^7$Be (spin $3/2^-$) into the ground state (spin $3/2^-$,
$E_{\nu_e}=0.863~MeV$, BR=90\%) and the excited state (spin $1/2^-$,
$E_{\nu_e}=0.381~MeV$, BR=10\%) of $^7$Li. Since solar neutrinos with
$E_{\nu_e}=0.381~MeV$ are below the threshold (0.81 MeV) for capture in
$^{37}$Cl and have a capture cross section in $^{71}$Ga that is smaller by
about a factor of 6 relative to solar neutrinos with
$E_{\nu_e}=0.863~MeV$, therefore a large suppression in the branching
ratio to the ground state can produce large suppressions of the $^7$Be
solar neutrino signals in $^{37}$Cl and in $^{71}$Ga. However, such an
explanation requires anomalously large fields near the center of the sun. 
 
\noindent {\bf Temporal and Spatial Variations in T:} Davis (1996) has
been claiming persistently that the solar neutrino flux measured by him
and his collaborators in the $^{37}$Cl radiochemical experiment is varying
with time. Because of the possibility that neutrinos may have anomalous
magnetic moments, much larger than those predicted by minimal extensions
of the standard electroweak model, which can solve the solar neutrino
problem, attention has been focused on anticorrelation between the solar
magnetic activity (the 11 year cycle) and the $\nu_\odot$ flux (see, e.g.,
Davis 1996). Also a day-night effect (e.g., Cribier et al 1986; Dar and
Mann 1987) due to resonant conversion of the lepton flavor of solar
neutrinos which cross Earth at night before reaching the solar neutrino
detector was not found by Kamiokande. However, the basic general question
whether the solar neutrino flux varies on a short time scale, has not been
fully answered, mainly because of the limited statistics of the first
generation of solar neutrino experiments.  \noindent The SSM predict no
significant variation of the solar neutrino flux on time scales shorter
than millions of years. However, the sun has a differential rotation. It
rotates once in $\sim$ 25 days near the equator, and in $\sim$ 33 days
near the poles. Moreover, the observed surface rotation rates of young
solar-type stars are up to 50 times that of the sun. It suggest that the
sun has been losing angular momentum over its lifetime. The overall
spin-down of a sun-like star by mass loss and electromagnetic radiation is
difficult to estimate from stellar evolution theory, because it depends on
delicate balance between circulations and instabilities that tend to mix
the interior and magnetic fields that retard or modify such processes. It
is quite possible that the differential rotation extends deep into the
core of the sun and causes there spatial and temporal variations in the
solar properties due to circulation, turbulences and mixing. Since R is
very sensitive to the temperature, even small variations in temperature
can affect R significantly without affecting significantly the pp solar
neutrino flux (the $^7$Be and $^8$B solar neutrinos will come mainly from
temperature peaks, while the pp neutrinos will reflect more the average
temperature).  \noindent If the solar neutrino flux is time dependent,
then cross correlation analysis of the various data sets from the
Homestake, Kamiokande, GALLEX, SAGE and Superkamiokande may reveal such
unexpected correlations: If arbitrary time lags are added to the different
solar neutrino experiments, the cross correlation is maximal when these
time lags vanish.  Moreover, a power spectrum analysis of the signals may
show peaks, if the time variation is periodic. In particular,
Super-Kamiokande with its high statistics should examine whether data from
different fiducial volumes are cross correlated in time.  Relevant
information on time variability in the solar core may come soon also from
SOHO and GONG. 

\noindent 
{\bf Mixing of $^3$He:} 
The SSM $^3$He equilibrium abundance increases sharply with radius.
Cummings and Haxton (1996) have recently suggested that the $^7$Be solar
neutrino problem could be circumvented in models where $^3$He is
transported into the core in a mixing pattern involving rapid filamental
flow downward. We note that if this mixing produces hot spots (due to
enhanced energy release) they can increase the effective temperature for p
capture by $^7$Be in a cool environment, reducing R while keeping the
$^8$B solar neutrino flux at the observed level. Perhaps, helioseismology
will be able to test that. 

\noindent
Cummings and Haxton (1996) also noted that such mixing will have other
astrophysical consequences. For example, galactic evolution models predict
$^3$He abundances in the presolar nebula and in the present interstellar
medium (ISM) that are substantially (i.e., a factor of five or more) in
excess of the observationally inferred values.  This enrichment of the ISM
is driven by low-mass stars in the red giant phase, when the convective
envelope reaches a sufficient depth to mix the $^3$He peak, established
during the main sequence, over the outer portions of the star.  The $^3$He
is then carried into the ISM by the red giant wind. The core mixing lowers
the main sequence $^3$He abundance at large r.

\section{Conclusions} 
The solar neutrino problem does not provide conclusive evidence for 
neutrino properties beyond the standard electroweak model.
The solar neutrino problem may be an astrophysical problem. 
The deviations of the experimental results from those predicted by
the standard solar models may reflect the approximate nature of 
our knowledge of nuclear reaction rates and  radiation transport  
in dense stellar plasmas and the approximate nature of the solar models
which neglect angular momentum effects, differential rotation, magnetic
field, angular momentum loss and mass loss during evolution and do not
explain yet, e.g., solar activity and the surface depletion of lithium,
relative to its meteoritic value (which may or may
not be relevant to the solar neutrino problem). 
Improvements of the
standard solar model should continue. In particular, dense plasma effects
(screening, correlations, fluctuations and deviations from Maxwell-
Boltzmann distributions) on nuclear reaction rates and radiative
opacities, which are not well understood, may affect the SSM predictions
and should be further studied, both theoretically and experimentally.
Relevant information may be obtained from studies of thermonuclear plasmas
in inertial confinement experiments. Useful information may also be
obtained from improved data on screening effects in low energy nuclear
cross sections of ions, atomic beams and molecular beams incident on a
variety of gas, solid and plasma targets. 

Better knowledge of low energy nuclear cross sections is still
needed. Improved measurement of 
the low energy nuclear cross sections  for the radiative
captures ${\rm p+^7Be\rightarrow ^8B+\gamma}$ and ${\rm
^3He+^4He\rightarrow ^7Be+\gamma}$ by photodissociation of $^8$B and
$^7$Be in the Coulomb field of heavy nuclei can help determine whether
there is a $^8$B solar neutrino problem. 
 
\noindent

Neutrino oscillations, and in particular the MSW effect, may be the
correct solution to the solar neutrino problem. But, only future
experiments, such as SNO, BOREXINO and HELLAZ, will be able to supply a
definite proof that Nature has made use of this beautiful effect. 

\noindent
{\bf Acknowledgement:} This work 
was supported in part by the Technion fund for the
promotion of research and by the N. Harr and R. Zinn research fund. 

\vfill
\eject
\bigskip
\section*{References}

$~~~~$Abdurashitov, J. N., et al., (SAGE), 1996, Phys. Rev. Lett. {\bf 77}, 
4708\\

Adelberger, E., et al., 1998,  preprint astro-ph/9805121\\ 

Akhmedov, E. Kh., 1988, Phys. Lett. B {\bf 213}, 64\\ 

Alexander, T. K., et al., 1984 Nucl. Phys. A {\bf 427}, 526 \\

Bahcall, J. N., 1989, {\it Neutrino Astrophysics} (Cambridge
Univ. Press 1989).\\
 
Bahcall, J. N. \& Bethe, H., 1991, Phys. Rev. D. {\bf 44}, 2962\\
 
Bahcall, J. N. \& Pinsonneault, M. H.,  1995, Rev. Mod. Phys. {\bf 67}, 781\\

Bahcall, J. N., Basu, S.  \& Pinsonneault, M. H., 1998, Astro-ph/9805135\\

Barker, F. C., 1995, Nucl. Phys. A {\bf 588}, 693\\  

Basu, S. \& Antia, H. M., 1995, MNRAS {\bf 276}, 1402 \\

Basu, S. \& Antia, H. M., 1997, MNRAS {\bf 287}, 189 \\

Brun, A. S., Turck-Chieze, S. \& Morel, P., 1998, ApJ. {\bf 506}, p. \\      

Clayton, D. 1968, {\it Princ. of Stellar Evolution \& Nucleosyn.}
(McGraw-Hill)\\

Cleveland, B.T., et al. (Homestake), 1998, ApJ. {\bf 496}, 505\\ 

Christensen-Dalsgaard, J., 1996, Nucl. Phys. B (Proc. Suppl.) {\bf 48}, 325\\

Cribier, M. et al., 1986, Phys. Lett. {\bf B182 2}, 89\\

Cummings, A. \& Haxton, W., 1996, preprint nucl-th 9608045\\

Cunha, K.  \& Lambert, D. L., 1992 ApJ, {\bf 399}, 586\\  
 
Dar, A. \& Mann, A., 1987, Nature {\bf 325}, 790\\
 
Dar, A.,  et al., 1987, Phys. Rev. D, {\bf 35}, 3607\\

Dar, A. \& Nussinov, S., 1991, Particle World {\bf 2}, 117\\

Dar, A., 1993, {\it Particles and Cosmology} p. 3 (World Scientific,
eds. E. N.\\$~~~~$ Alexeev et al.)\\ 

Dar, A. \& Shaviv, G., 1996, ApJ, {\bf 468}, 933 \\ 
 
Davis, R. Jr. 1996, Nucl. Phys. B (Proc. Suppl.) {\bf 48}, 284\\

Dwarakanath, M. R., \& Winkler, H., 1971, Phys. Rev. C. {\bf 4}, 1532 
 
Filippone, B., et al., 1983a, Phys. Rev. Lett. {\bf 50}, 412\\

Filippone, B., et al., 1983b, Phys. Rev. C {\bf 28}, 2222 \\

Fukuda, Y., et al. (Kamiokande), 1996, Phys. Rev. Lett. {\bf 77}, 1683\\ 

Fukuda, Y., et al., 1998, submitted to Phys. Rev. Lett. hep-ph/9805021\\ 

Gies, D. R. \& Lambert, D. L., 1992 ApJ, {\bf 387}, 673\\  

Gavrin, V., et al. (SAGE), 1998, {\it Neutrino 98}, Nucl. Phys. B (Proc. 
Suppl) \\ 

Grevesse, N.,  Noels, A., \& Sauval, A. J., 
1996, in {\it Cosmic Abundances},\\
(eds. S. S. Holt \& G. Sonneborn), APS Conference Series, {\bf 99}, 117 \\

Greife, U., et al., 1994, Nucl. Ins. \& Meth. in Phys. Res. A
{\bf 350}, 327\\

Hampel, W. et al. (GALLEX), 1996, Phys. Lett. B {\bf388}, 384\\   

Hammache, F., et al. (LUNA), 1998, Phys. Rev. Lett. {\bf 80}, 928 \\

Hilgemeier, M., et al., 1988 , Z. Phys. A. {\bf 329}, 243\\

Hill, F. et al. (GOLF), 1996, Science {\bf 272}, 1292\\  

Hernandez, E. P. \& Christensen-Dalsgaard, J., (1994), MNRAS
{\bf 269}, 475\\

Iglesias, C. \& Rogers, F.J., 1996, ApJ, {\bf 464}, 943  \\ 

Junker, M., et al., (LUNA) , preprint nucl-ex/

Kamionkowski, M. \& Bahcall, J. N., 1994, ApJ, {\bf 420}, 884\\

Kaniadakis, G. et al., 1998, preprint astro-ph/9710173 \\   

Kavanagh, R. W., 1960 Nucl. Phys. {\bf 15}, 14\\

Kavanagh, R. W., et al.,  1969, Bull. Am. Nucl. Phys. Soc.
{\bf 14}, 1209\\

Kirsten, T., et al., 1998 {\it Neutrino 98}, Nucl. Phys. B (Proc. Suppl.\\ 
   
Kovetz, A. \& Shaviv, G., 1994, ApJ. {\bf 426}, 787\\
 
Krauss, A., et al., 1987, Nucl. Phys. A. {\bf 467}, 273\\ 

Kr\"awinkel, H., et al., 1982, Z. Phys. A {\bf 304}, 307 \\

Lim, C. S. \& Marciano, W. J., 1988, Phys. Rev. D {\bf37}, 1368 \\

Mikheyev, P. \& Smirnov, A. Yu. 1985, Yad. Fiz. {\bf 42}, 1441\\

Motobayashi, T., et al., 1994, Phys. Rev. Lett. {\bf 73}, 2680 \\

Nagatani, et al., 1969, Nucl. Phys.\\  

Nunes, F. M.,  et al.,  1997, Nucl. Phys. A {\bf 615}, 69\\

Okun, L. B., et al., 1986; Sov. J. Nucl. Phys. {\bf 44}, 440\\

Osborne, J. L., et al., 1982, Phys. Rev. Lett. {\bf 48}, 1664\\

Osborne, J. L., et al., 1984, Nucl. Phys. A {\bf 419}, 115\\

Parker, P. D. \& Kavanagh, R. W. 1963, Phys. Rev. {\bf 131}, 2578\\

Parker, P. D., 1966, Phys. Rev.  {\bf 130}, 851\\

Parker, P. D., 1968, ApJ. Lett.  {\bf 153}, 85\\

Richard, O., et al., 1996, A \& A, {\bf 312}, 1000 \\

Robertson, R. G. H., et al., 1983, Phys. Rev. C {\bf 27}, 11 \\

Rogers, F.J., et al., 1996, ApJ, {\bf 456}, 902\\

Salpeter, E. E. \& Van Horne, H. M.., 1969, ApJ, {\bf 155}, 183  \\

Shaviv, G. \& Shaviv, N., 1996, ApJ, {\bf 468}, 433\\

Shaviv, G. \& Shaviv, N., 1998, in preparation\\

Suzuki, Y., et al., 1998, {\it Neutrino 98}, Nucl. Phys. B (Proc. Suppl) \\ 

Turck-Chieze, S., et al., 1997, Sol. Phys. {\bf 175}, 247\\

Vaughn, F. J., et al., 1970, Phys. Rev. C {\bf 2}, 1657 \\

Volk, H., et al., 1983, Z. Phys. A {\bf 310}, 91\\

Wilson, T. L. \& Rood, R. T., 1994, Ann. Rev. Astr. Ap. {\bf 32}, 1\\

Wolfenstein, L., 1978, Phys. Rev. {\bf D17}, 2369\\
 
Wolfenstein, L., 1979, Phys. Rev. D {\bf 20}, 2634\\

Xu, H. M., et al., 1994, Phys. Rev. Lett. {\bf 73}, 2027\\

\clearpage 
 
\noindent
{\bf Table Ia:} Comparison between the solar neutrino fluxes predicted
by the SSM of BP95, BSP98, BTM98 and   
DS96, and measured by the solar neutrino experiments.
$$\matrix{~~~~~\nu~{\rm Flux}\hfill 
& {\rm BP95} \hfill & {\rm BSP98} \hfill 
& {\rm BSP98}\hfill & {\rm   DS96}\hfill 
&{\rm Experiment} \hfill \cr
{\rm \phi }_{\nu }(pp)~[{10}^{10}{cm}^{-2}{s}^{-1}] \hfill 
&5.91\hfill &5.94\hfill &{\rm \hfill} &6.10 \hfill  &\rm \cr
{\phi }_{\nu }(pep)~[{10}^{8}{cm}^{-2}{s}^{-1}] \hfill 
&1.39\hfill &1.39\hfill &{\rm \hfill} &1.43\hfill &\rm \hfill\cr
{\phi }_{\nu }(^{7}{Be})~[{10}^{9}{cm}^{-2}{s}^{-1}]\hfill 
&5.18\hfill&4.80\hfill &{\rm \hfill}&3.71\hfill &\rm \hfill \cr
{\phi }_{\nu }(^{8}{B})~[{10}^{6}{cm}^{-2}{s}^{-1}]\hfill 
&6.48\hfill &5.15\hfill &4.82\hfill
& 2.49 \hfill &\rm 2.44\pm 0.11\hfill\cr
{\phi }_{\nu }(^{13}{N})~[{10}^{8}{cm}^{-2}{s}^{-1}]\hfill 
&6.4\hfill&6.05\hfill& \hfill & 3.82 \hfill &\rm \hfill \cr
{\phi }_{\nu }(^{15}{O})~[{10}^{8}{cm}^{-2}{s}^{-1}]\hfill 
&5.15\hfill &5.32\hfill&\hfill & 3.74 \hfill &\rm \hfill \cr
{\phi }_{\nu }(^{17}{F})~[{10}^{6}{cm}^{-2}{s}^{-1}]\hfill 
&6.48\hfill&6.33\hfill& \hfill&4.53\hfill & \hfill\cr
{\rm Rates}\hfill& \hfill & \hfill &\hfill &\hfill&\hfill\cr
\Sigma (\phi \sigma)_{Cl}~[SNU]\hfill&  
9.3\pm 1.4\hfill &7.7\pm 1.2\hfill &7.18\hfill 
 &4.1\pm 1.2\hfill & 2.56\pm 0.25\hfill \cr
\Sigma(\phi\sigma)_{Ga}~[SNU]\hfill 
& 137\pm 8\hfill& 129\pm 8\hfill& 127\pm 8\hfill
& 115\pm 6\hfill&76.4\pm 8 \hfill\cr
\Sigma(\phi\sigma)_{Ga}~[SNU]\hfill 
& 137\pm 8\hfill& 129\pm 8\hfill& 127\pm 8\hfill& 115\pm 6\hfill
&70\pm 8 \hfill\cr}$$

\noindent
{\bf Table Ib} Characteristics of the BP95, BTM98, and DS96
Solar Models in Table Ia
(c=center; s=surface; bc=base of convective zone;
${\rm \bar N=log([N]/[H])+12)}$.
 
$$\matrix{{\rm Parameter}\hfill& {\rm BP95}\hfill& {\rm
BTM98}\hfill& {\rm
DS96}\hfill \cr
{T}_{c}~[{10}^{7}K] \hfill &1.584 \hfill &1.567 \hfill &1.561 \hfill \cr
{\rho }_{c}~[g~c{m}^{-3}]\hfill&156.2 \hfill&151.9\hfill&155.4\hfill
\cr
{X}_{c}\hfill&0.3333\hfill &0.3442\hfill &0.3424 \hfill \cr
{Y}_{c}\hfill&0.6456 \hfill &0.635 \hfill &0.6380 \hfill \cr
{Z}_{c}\hfill&0.0211\hfill&0.02084 \hfill
&0.01940 \hfill \cr
{R}_{conv}~[R/R_{\odot}]\hfill&0.712 \hfill &0.715 \hfill &0.7130
\hfill \cr
{T}_{bc}~[{10}^{6}{\rm K}]\hfill&2.20 \hfill &2.172 \hfill &2.105
\hfill \cr
{X}_{s}\hfill&0.7351 \hfill &0.739 \hfill &0.7512 \hfill \cr
{Y}_{s}\hfill&0.2470 \hfill &0.243 \hfill &0.2308 \hfill \cr
{Z}_{s}\hfill&0.01798 \hfill &0.0181 \hfill
&0.0170 \hfill \cr
\overline{N}_s{(^{12}C})\hfill&8.55\hfill &8.55 \hfill&8.55 \hfill \cr
\overline{N}_s{(^{14}N})\hfill&7.97 \hfill &7.97 \hfill&7.97\hfill \cr
\overline{N}_s{(^{16}O})\hfill&8.87 \hfill &8.87 \hfill&8.87 \hfill \cr
\overline{N}_s{(^{20}Ne})\hfill&8.08 \hfill &8.08 \hfill&8.08 \hfill\cr 
{T}_{eff}~[{\rm K}]\hfill& 5800\hfill &5800 \hfill &5803 \hfill \cr}$$
 
\clearpage 
 
\noindent
{\bf Table II:} Comparison between the SSM of Bahcall and Pinsonneult
(1995) and of Dar and Shaviv (1996).
$$\matrix{{\rm Parameter}  
\hfill& {\rm ~~~~~~BSP98}\hfill& {\rm 
~~~~~~DS96}\hfill \cr
\hfill&\hfill& \hfill \cr
{M}_{\odot} \hfill &1.9899\times 10^{33}~g \hfill &1.9899\times 10^{33}~g 
\hfill\cr
{L}_{\odot} \hfill&3.844\times 10^{33}~erg~s^{-1} \hfill& 3.844\times 10^{33}
~erg~s^{-1}\hfill\cr
{R}_{\odot}\hfill&6.9599\times 10^{10}~cm \hfill &6.9599\times 
10^{10}~cm\hfill \cr 
{t}_{\odot}\hfill&4.566\times 10^9~ y \hfill &4.57\times 10^9~y 
\hfill \cr 
{\rm Rotation}\hfill&{\rm Not~Included} \hfill & {\rm Not~Included} 
\hfill \cr 
{\rm Magnetic~Field}\hfill&{\rm Not~Included} \hfill & {\rm Not~Included} 
\hfill \cr 
{\rm Mass ~Loss}\hfill&{\rm Not~Included} \hfill & {\rm Not~Included} 
\hfill \cr 
{\rm Angular~Momentum~Loss}\hfill&{\rm Not~Included} \hfill & {\rm Not 
~Included} \hfill \cr 
{\rm Premain~Sequence~Evolution}\hfill&{\rm Not~ 
Included} \hfill & {\rm Included} \hfill \cr 
{\rm {\bf Initial~Abundances: }}\hfill& \hfill & \hfill \cr
{\rm ^4He}\hfill& {\rm Adjusted} \hfill &{\rm 
Adjusted} \hfill \cr 
{\rm C,N,O,Ne}\hfill& {\rm Adjusted} \hfill &{\rm 
Adjusted} \hfill \cr 
{\rm All~Other~Elements}\hfill& {\rm Adjusted} \hfill &{\rm 
Meteoritic} \hfill \cr 
{\rm {\bf Photospheric~Abundances: }}\hfill&\hfill & \hfill \cr
{\rm ^4He }\hfill&{\rm Predicted} \hfill&{\rm Predicted}\hfill \cr
{\rm C,N,O,Ne}\hfill& {\rm Photospheric} \hfill &{\rm Photospheric} \hfill 
\cr {\rm All~Other~Elements}\hfill& {\rm Meteoritic} \hfill &{\rm 
Predicted} \hfill \cr 
\hfill& \hfill & \hfill \cr
{\rm Radiative~Opacities}\hfill&{\rm OPAL~1996}
\hfill & {\rm OPAL~1996} \hfill \cr 
{\rm Equation~ of~ State}\hfill&{\rm Straniero ~1996?}
\hfill & {\rm DS~ 1996} \hfill \cr 
{\rm Partial~ Ionization~ Effects}\hfill&{\rm Not~ 
Included} \hfill & {\rm Included} \hfill \cr 
{\rm Diffusion~ of~ Elements:}\hfill& \hfill & \hfill \cr
{\rm H,~^4He}\hfill&{\rm Included} \hfill&{\rm Included} \hfill \cr
{\rm Heavier~Elements}\hfill&{\rm Approximated~ by~ Fe } \hfill&{\rm 
 All~Included} \hfill \cr
{\rm Partial~ Ionization~ Effects}\hfill&{\rm Not~Included } \hfill&{\rm 
 Included} \hfill \cr
{\rm Nuclear~ Reaction~ Rates:}\hfill& \hfill & \hfill \cr 
\rm S_{11}(0)~eV~\cdot b\hfill &
\rm 4.00\times {10}^{-19}\hfill &\rm 4.07\times {10}^{-19}\hfill\cr
\rm S_{33}(0)~MeV\cdot b \hfill &
5.3\times \hfill &5.6 \hfill \cr 
\rm S_{34}(0)~ keV\cdot b\hfill &0.53 \hfill & 0.45\hfill \cr 
\rm S_{17}(0)~eV\cdot b\hfill&19\hfill
& 17\hfill \cr
{\rm Screening~ Effects}\hfill &{\rm Included } \hfill&{\rm Included} 
\hfill \cr 
{\rm Nuclear~Equilibrium}\hfill&{\rm Imposed} \hfill&{\rm Not~Assumed} 
\hfill \cr}$$
\clearpage

\clearpage
 
\noindent
{\bf Table IV:} Fractional change in the predicted $\nu_\odot$ fluxes 
and counting rates in the $\nu_\odot$ experiments due to the inclusion of 
element diffusion in the SSM calculations of Bahcall and Pinsonneault 
(BP95), Brun, Turck-Chieze and Morel (BTM98), Dar and Shaviv 
(DS96) and Richard, Vauclair, Charbonnel and Dziembowski (RVCD96).
The results of models 1 and 2 of RVCD96 were   
extrapolated to the initial solar composition which was used in DS96. 
 
$$\matrix{\phi_{\nu_\odot} & {\rm BP95}&{\rm BTM98}& {\rm DS96} & {\rm 
RVCD96}\cr pp\hfill &-~1.7\%\hfill&\hfill &-~0.3\%\hfill &-~0.8\% \hfill\cr
pep\hfill &-~2.8\%\hfill&\hfill &-~0.3\%\hfill &-~0.4\%\hfill\cr
{\rm ^7Be}\hfill    &+13.7\%\hfill&\hfill &+4.2\%\hfill  &+ ~6.5\%\hfill\cr
{\rm ^8B}\hfill  &+36.5\%\hfill &+32~~\%\hfill &+11.2\%\hfill 
&+10.7\%\hfill\cr
{\rm ^{13}N}\hfill &+51.8\%\hfill &\hfill &+22.7\%\hfill &+19.8\%\hfill\cr
{\rm ^{15}O}\hfill &+58.0\%\hfill&\hfill &+24.0\%\hfill &+20.8\%\hfill\cr
{\rm ^{17}F}\hfill &+61.2\%\hfill &\hfill &+24.9\%\hfill &+21.8\%\hfill\cr
{\rm Rates} & & & \cr
{\rm H2O}\hfill &+36.5\%\hfill &+32~~\%\hfill &+11.2\%\hfill&+13.3\%\hfill\cr
{\rm Cl}\hfill &+32.9\%\hfill &+27~~\%\hfill &+~9.5\%\hfill&+12.3\%\hfill\cr
{\rm Ga}\hfill &+~8.7\%\hfill &+~7~~\%\hfill 
&+~2.6\%\hfill&+~3.7\%\hfill\cr}$$
 
\vfill
\eject
\section*{Figure Captions}

\noindent 
{\bf Fig. 1.} The principal branches of the pp cycle and the CNO bi-cycle.\\

\noindent 
{\bf Fig. 2.} The solar neutrino capture rate in $^{37}$Cl as measured
in the Homestake experiment runs nos. 18 t0 133. The dashed line 
shows the average value.\\

\noindent 
{\bf
Fig. 3.} The $^8$B solar neutrino flux as function of time as measured
by Kamiokande between 1986-1996.\\

\noindent 
{\bf Fig. 4.} The cosine of the angle between the electron direction
and the radius vector from the sun in Super-Kamiokande.
The solid line shows the best fit for a $^8$B
solar neutrino flux.\\

\noindent 
{\bf Fig. 5.} The time variation of the $^8$B solar neutrino flux
as measured by Super-Kamiokande from June 96 to June 97.
The solid line shows the expected variation of the flux due to the
eccentricity of the Earth's orbit around the sun.\\

\noindent
{\bf Fig. 6.} The ratio between the $^8$B solar neutrino flux observed
by Super-Kamiokande and the flux predicted by the SSM of BP95
as function of zenith angle of the sun. \\

\noindent
{\bf Fig. 7.} The ratio between the observed number of electrons
scattered by solar neutrinos in Super-Kamiokande and their expected 
number in the  SSM of BP95 as function of electron recoil energy.\\

\noindent
{\bf Fig. 10} The $^{71}$Ge production rate in $^{71}$Ga by solar 
neutrinos as measured by GALLEX and SAGE between 1990 and 1997.\\

\noindent
{\bf Fig. 9.} The astrophysical S(E) factor for the reaction
${\rm ^3He+^3He\rightarrow ^4He+2p}$ as measured by various
low energy experiments. The dotted line is a best  
polynomial fit (solid line) to the data with maximal screening 
enhancement ($U_e=240~eV$).\\  

\noindent
{\bf Fig. 10.} The relative difference beteen the speed of sound squared as 
inferred from the helioseismological measurements of GOLF and LOWL
on board SOHO and that calculated in BTM98 from
the SSM with diffusion (solid line) and without diffusion (dotted line).\\ 

\end{document}